\documentclass[final,3p,times,authoryear]{elsarticle}
\pdfoutput=1
\usepackage{natbib}
\usepackage{graphicx}
\usepackage{amsmath,amsfonts, amsthm, mathrsfs, amssymb, bbm}
\usepackage{times}

\begin{document}

\begin{frontmatter}

\title{Exact solution of a two-type branching process:\\
Clone size distribution in cell division kinetics}
\author[PED]{Tibor Antal}
\ead{tibor\_antal@harvard.edu}
\author[BU]{P. L. Krapivsky}
\ead{paulk@bu.edu}
\address[PED]{Program for Evolutionary Dynamics, Harvard University,
  Cambridge, MA~ 02138, USA}
\address[BU]{Department of Physics, Boston University, Boston, MA~ 02215, USA}

\begin{abstract}
We study a two-type branching process which provides excellent description of experimental data on cell dynamics in skin tissue \citep{clayton07}. The model involves only a single type of progenitor cell, and does not require support from a self-renewed population of stem cells. The progenitor cells divide and may differentiate into post-mitotic cells. We derive an exact solution of this model in terms of generating functions for the total number of cells, and for the number of cells of different types. We also deduce large time asymptotic behaviors drawing on our exact results, and on an independent diffusion approximation. 
\end{abstract}


\end{frontmatter}

\section{Introduction}

Understanding the kinetics (homeostasis) of cells in adult mammalian tissues has long been a major challenge in biology. Recent progress in experimental methods made it feasible to label individual cells {\it in vivo}, and follow their fate and that of their progeny \citep{clarke99,jonkers02}. This powerful genetic labeling technique has enabled {\it in vivo} experiments in the outmost layer of skin (epidermis) of the tail in adult mice \citep{clayton07}. Individual cells in the basal layer of the epidermis have been marked by a fluorescent genetic label and the size of the clone (all living progenies of a cell) of each single marked cell has been measured at different times. This has provided the data on the evolution of the clone size distribution in the basal layer of the epidermis. 

The prevailing model of epidermal homeostasis has involved long-lived stem cells generating short-lived populations of transit-amplifying (TA) cells that differentiate into post-mitotic cells \citep{potten74,elaine09}. The stem-TA hypothesis predicts that the clones of TA cells should disappear (after sufficiently long time), while the existing clones should be small and associated with stem cells. Strikingly, the fraction of remaining clones was found to decrease as $({\rm time})^{-1}$; accordingly the average size of existing clones scales linearly with time. This remarkable scaling behavior calls for a totally different model of epidermal homeostasis. \citet{clayton07} proposed a model of cell division and differentiation which manifestly obeys the observed scaling behavior and provides an excellent fit to more subtle characteristics. A gratifying property of the model suggested by \citet{clayton07} is that it is {\em simpler} than the stem-TA model: The new model involves only a single type of committed progenitor cell, and in particular, stem-cell proliferation is not required for epidermal homeostasis. 

Thus the model describes the population of cells of two types: Proliferating cells (type $A$) divide and eventually differentiate into non-proliferiting cells (type $B$), which leave the basal layer and migrate to the epidermal surface where they are shed. More precisely, the cell population evolves according to the continuous time, constant rate, two-type branching process
\begin{equation}
\label{modeldef}
 \begin{tabular}{ll}
 $A \to  A A$  ~~~ & at rate $r$\\
 $A \to  A B$ & at rate $1-2r$\\
 $A \to  B B$ & at rate $r$\\
 $B \to  \emptyset$ & at rate $\gamma$\\
 \end{tabular}
\end{equation}
Here we set the overall cell division rate to unity. In the experiments of \citet{clayton07},
the division rate was equal to $\lambda=1.1/$week; the values of the parameters were found to be $r=0.08$ and $\gamma\equiv\Gamma/\lambda=0.28$. 
Note that the model is assumed to be critical, that is the division rates corresponding to the channels  $A \to  A A$ and $A \to  B B$ are the same. Due to this symmetry, the average population size of progenitor cells remains constant as it is required by the steady-state assumption. The average population size of the post-mitotic cells is also constant. 

So far the model has been experimentally tested only in mice tail skin. There are still technical constraints preventing the quantitative tests of the model in other tissues, but  those problems are temporary. The model challenges the necessity of stem-cell proliferation for the homeostasis of epidermis \citep{JS08}.  There is also growing evidence \citep{dor04,adam09} that stem cells do not contribute to the maintenance of various other adult tissues. Hence the two-type branching process \eqref{modeldef} may find a broad range of applications and therefore it is highly desirable to possess an exact solution. Despite its apparent simplicity, the branching process \eqref{modeldef} has not been solved, although some exact and asymptotic behaviors have been found \citep{clayton07,klein07,klein08}. In this paper we apply generating function techniques to obtain an exact analytic solution, as well as approximate methods to derive asymptotic limits. 

Branching processes have been extensively used to model proliferation of differentiating cells, especially in the hemopoietic (blood production) system \citep{vogel69,pharr85}; see also other references in Section 6.9.1 in \citep{kimmel02}. An interesting multi-type model has also been proposed recently in \citep{dingli07b,dingli09}.
These studies, however, mainly rely on numerical solutions, while analytic treatment is restricted to obtaining average quantities (or second moments).

The rest of the paper is organized as follows. We introduce the model in Section~\ref{model}, and discuss its basic behavior. After presenting the generating function methods in Section~\ref{generate}, we provide an elementary solution on a special line in the parameter space in Section~\ref{elementary}. The model admits a neat explicit solution at the special point $\gamma=1, r=1/4$, which is discussed in Section~\ref{special}. As our main result, we derive the generating function of the model for general parameter values in Section~\ref{general}, where we also present an efficient numerical method to obtain the probabilities of having certain number of cells at a given time. We discuss the large time asymptotic behavior in Section~\ref{scaling}, and derive additional scaling properties by means of the Fokker-Plank method in Section~\ref{diffuz}. Final remarks are presented in Section~\ref{disco}.

\section{The Model}
\label{model}

The model involves two types of cells, $A$ and $B$. Type $A$ cells (progenitor cells) are able to divide (proliferate) and diffirentiate, $B$ cells (post-mitotic cells) do not divide, they just die (leave the basal layer). More precisely, the two cell populations evolve according to the two-type branching process \eqref{modeldef}. The probability $P_{m,n}(t)$ of having $m$ copies of $A$, and $n$ copies of $B$ at time $t$ satisfies
\begin{equation}
\label{master}
 \frac{d P_{m,n}}{dt} = r(m-1)P_{m-1,n} + (1-2r)mP_{m,n-1} + r(m+1)P_{m+1,n-2} + \gamma (n+1)P_{m,n+1}
   - (m+\gamma n)P_{m,n}
\end{equation}
The consecutive gain terms on the right-hand side of Eq.~\eqref{master} merely describe the contributions of the channels (from top to bottom) of the two-type branching process \eqref{modeldef}. 
To determine the clone size distribution we start with a single $A$ cell, that is 
\begin{equation}
\label{initial}
P_{m,n}(t=0)=\delta_{m,1}\delta_{n,0}
\end{equation}
We are interested in the full distribution $P_{m,n}(t)$ and also in the reduced probability distribution $\Pi_s(t)$ of having $s=m+n$ total cells at time $t$; the latter distribution is directly probed in experiments. Needless to say, 
\begin{equation}
\Pi_s(t) = \sum_{m+n=s} P_{m,n}(t)
\end{equation}

Let us first determine the probability distribution $P_m(t)$ of having $m$ cells of type $A$.  This probability distribution is readily found since $B$ cells do not affect $A$ cells, and $A$ cells alone evolve according to the critical branching process
\begin{equation}
 \begin{tabular}{ll}
 $A \to  A A$  ~~~ & at rate $r$\\
 $A \to  \emptyset$ & at rate $r$\\
 \end{tabular}
\end{equation}
The solution, for the initial condition $P_{m}(t=0)=\delta_{m,1}$, is \citep{athreya04}
\begin{equation}
\label{Pmt}
P_m(t) = \left\{
\begin{tabular}{ll}
$\displaystyle \frac{1}{(1+rt)^2} \left( \frac{rt}{1+rt} \right)^{m-1}$ & for $m\ge 1$\\
$\displaystyle \frac{rt}{1+rt}$ & for $m=0$
\end{tabular}
\right.
\end{equation}
Notice that the average number of $A$ cells remains constant, 
\begin{equation}
\langle m\rangle = \sum_{m\geq 0}mP_m(t) = 1,
\end{equation}
throughout the evolution. This is of course a general property of the critical branching process. 

We can also compute the average number of post-mitotic cells
$\langle n\rangle = \sum_{m, n\geq 0}nP_{m,n}(t)$. Indeed, this quantity satisfies a simple rate equation
\begin{equation}
\label{n-av}
\frac{d \langle n\rangle}{d t} = \langle m\rangle - \gamma \langle n\rangle
\end{equation}
The gain term on the right-hand side of \eqref{n-av} follows from the second and third channels (from top to bottom) of the two-type branching process \eqref{modeldef}; the loss term corresponds to the last channel. Using 
$\langle m\rangle = 1$ and $\langle n\rangle|_{t=0}=0$ we solve \eqref{n-av}  to yield 
\begin{equation}
\langle n\rangle =  \frac{1-e^{-\gamma t}}{\gamma}
\end{equation}
Therefore the total average number of cells is given by 
\begin{equation}
\langle s\rangle = 1 + \frac{1-e^{-\gamma t}}{\gamma}
\end{equation}
Note that the fraction of type $A$ cells is asymptotically
\begin{equation}
\label{rho}
 \rho\equiv \frac{\langle m\rangle}{\langle s\rangle} = \frac{\gamma}{1+\gamma}
\end{equation}

These exact expressions for the average population sizes are useful and e.g.\ the 
fraction of type $A$ cells \eqref{rho} will appear in numerous latter formulae. The full description of the clone size requires analyzing an infinite set of master equations \eqref{master}. We shall perform such analysis using generating function techniques.

\section{Generating function}
\label{generate}

We define the generating function of $P_{m,n}(t)$ as
\begin{equation}
\label{gener}
 F(x,y,t) = \sum_{m,n=0}^\infty x^m y^n P_{m,n}(t)
\end{equation}
Note that \eqref{master} is valid for all $m,n\ge 0$, if we define $P_{m,n}\equiv0$ for all $m<0$ or $n<0$. [Such systems are said to have {\it natural} boundary conditions \citep{kampen:1997aa}.] We multiply both sides of \eqref{master} by $x^m\,y^n$ and sum over all values of $m,n\ge 0$. Using identities $mx^m=x\partial_x x^m, ny^n
=y\partial_y y^n$, where $\partial_x=\partial/\partial x, \partial_y=\partial/\partial y$, we arrive at a partial differential equation 
\begin{equation}
\label{F-eq}
\partial_t F + \left[x(1-y)- r(x-y)^2\right]\partial_x F
 +\gamma (y-1)\, \partial_y F=0
\end{equation}
The initial condition \eqref{initial} corresponding to a single initial $A$ cell
becomes
\begin{equation}
\label{in-F}
F(x,y, t=0) = x
\end{equation}

Thus we need to solve the partial differential equation \eqref{F-eq} subject to \eqref{in-F}. Mathematically, equation \eqref{F-eq} is a hyperbolic partial differential equation and it can be analyzed using the method of characteristics \citep{Logan08}. Instead, we employ backward Kolmogorov equations; this approach is technically somewhat easier in the present case. Here we need two generating functions $F_A$ and $F_B$, where the subscripts refer to the type of the single initial cell. For the forward case we only needed the interesting $F\equiv F_A$. The initial conditions are
\begin{equation}
 \begin{split}
  F_A(x,y,t=0)&=x\\
  F_B(x,y,t=0)&=y
 \end{split}
\end{equation}
The coupled backward Kolmogorov equations read
\begin{subequations}
\begin{align}
  \label{backFA}
  \partial_t F_A &= rF_A^2 + (1-2r)F_A F_B +rF_B^2 -F_A\\
  \label{backFB}
  \partial_t F_B &= \gamma (1-F_B) 
\end{align}
\end{subequations}
These equations can be derived from the corresponding backward Kolmogorov equations for the probabilities $P_{m,n}(t)$, or they can be written down directly \citep{athreya04}. The negative terms ($-F_A$, $-\gamma F_B$) describe the disappearance of a cell, and the positive terms stand for the created new cells, with the corresponding rates. The term ``1" in \eqref{backFB} is just the generating function of no created particle, that is $1=x^0 y^0$.

Equation \eqref{backFB} is immediately solved to give
\begin{equation}
\label{onlyB}
 F_B = y e^{-\gamma t} +1-e^{-\gamma t}  \equiv f
\end{equation}
This is not surprising, of course: Starting with a single $B$ cell, the system will either contain the initial $B$ cell (this occurs with probability $e^{-\gamma t}$) or no cells at all. Substituting \eqref{onlyB} into \eqref{backFA} and changing the variable from $t$ to $f$ we obtain
\begin{equation}
 \gamma(1-f) \partial_f F = r(F-f)^2+F(f-1)
\end{equation}
where we dropped the subscript $A$ so that $F\equiv F_A$. We further simplify the 
above equation by changing variable $f$ to $u=1-f=(1-y)e^{-\gamma t}$. The function $F(u)$ then satisfies
\begin{equation}
\label{backmain}
 \gamma u F' = Fu-r(F+u-1)^2
\end{equation}
with initial condition $F(u=1-y)=x$. In equation \eqref{backmain} and later the prime denotes the derivative with respect to $u$. 
Note that the forward equation \eqref{F-eq} leads to the same equation  \eqref{backmain} via the method of characteristics.

Equation \eqref{backmain} is an ordinary differential equation of the first order. Yet it is non-linear and could be unsolvable as it belongs to the family of Riccati equations. Riccati equations are in principle intractable, yet there are two tricks which sometimes allow to solve certain Riccati equations \citep{Bender}. One is based on the reduction to the {\em linear} ordinary differential equation of the second order, the Sturm-Liouville equation. Another trick applies if we manage to find a special solution. We shall see that both tricks lead to success. Let us begin with the more elementary second approach.

\section{Elementary Solutions}
\label{elementary}

The idea is to guess one solution $F_*(u)$ irrespective whether it satisfies the initial condition or not.  Having found such a special solution, one then seeks a general solution in the form
\begin{equation}
\label{xx*}
F(u) = F_*(u) + \frac{1}{V(u)}
\end{equation}
The function $V(u)$ satisfies a {\em linear} differential equation which is readily solvable.

The form of \eqref{backmain} suggests to seek a special solution as a polynomial:
\begin{equation}
\label{x*}
F_*(u) = A_0 +A_1u+\ldots + A_p u^p
\end{equation}
Here $A_0,A_1,\ldots,A_p$ are constants and $A_p\ne 0$, so that the polynomial \eqref{x*} has degree $p$. Noting that $u F'$ is the polynomial of degree  $p$,
$u F$ is the polynomial of degree  $p+1$, and $(F-1+u)^2$ is the polynomial of degree  $2p$, equating the highest degree in $u$ would be possible only if $p+1=2p$, i.e. $p=1$. Thus the polynomial solution should be a linear function of $u$,
\begin{equation}
\label{x*-linear}
F_*(u) = A_0 +A_1u
\end{equation}
Plugging \eqref{x*-linear} into \eqref{backmain} we find that  the matching is achieved [that is, the ansatz \eqref{x*-linear} works] if $A_0=1, A_1=1/\gamma$ and the parameters $r, \gamma$ are related via
\begin{equation}
\label{rg-constraint}
r = \frac{\gamma}{(1+\gamma)^2}
\end{equation}

The prescription \eqref{xx*} tells us to seek the general solution in the form
\begin{equation}
\label{xx*-linear}
F = 1+\frac{u}{\gamma} + \frac{1}{V(u)}
\end{equation}
By inserting \eqref{xx*-linear} into \eqref{backmain} we arrive at a linear ordinary differential equation
\begin{equation}
\label{v-eq}
V' + \hat\gamma V - \frac{1}{(1+\gamma)^2}\,\frac{1}{u} = 0, \qquad 
\hat\gamma  = \frac{1}{\gamma}\,\frac{\gamma -1}{\gamma+1}
\end{equation}
The homogeneous part has solution $e^{-\hat\gamma u}$ and therefore the general solution to \eqref{v-eq} is sought as $V=e^{-\hat\gamma u}\,W$. The auxiliary function $W$ obeys
\begin{equation}
W' = \frac{1}{(1+\gamma)^2}\,\frac{e^{\hat\gamma u}}{u}
\end{equation}
which is solved to yield
\begin{equation}
\label{w-sol}
W = \frac{Ei(\hat\gamma u)}{(1+\gamma)^2} + {\rm const}
\end{equation}
Here $Ei(x)=-\int_{-x}^\infty d\xi\,e^{-\xi}/\xi$ is the exponential integral.
The constant (and the choice of the appropriate low limit in the integral) in \eqref{w-sol} are fixed by the initial condition. Recall that initially we have $F(u=1-y)=x$. Hence \eqref{xx*-linear} gives
\begin{equation}
x -1+\frac{y -1}{\gamma} = \frac{1}{V_0}
\end{equation}
and therefore
\begin{equation}
\label{w0}
W_0=V_0\,e^{\hat\gamma u_0}= V_0\,e^{\hat\gamma (1-y)} = 
\frac{e^{\hat\gamma (1-y)}}{x -1+(y -1)/\gamma}
\end{equation}
Using \eqref{w-sol} and \eqref{w0} we obtain 
\begin{equation}
\label{w-full}
W = \frac{e^{\hat\gamma (1-y)}}{x -1+(y -1)/\gamma} -\mathcal{E}(y,t)
\end{equation}
where we used the shorthand notation
\begin{equation}
\mathcal{E}(y,t) = 
\frac{Ei[\hat\gamma (1-y)] - Ei[\hat\gamma (1-y)e^{-\gamma t}]}{(1+\gamma)^2}
\end{equation}

Combining \eqref{xx*-linear} and \eqref{w-full} we arrive at 
\begin{equation}
\label{F-gen}
F(x,y,t) = 1+\gamma^{-1}\,e^{-\gamma t}\,(1-y)
+ e^{\hat\gamma  e^{-\gamma t}\,(1-y)}
\left[\frac{e^{\hat\gamma (1-y)}}{x -1+(y -1)/\gamma} - \mathcal{E}(y,t)\right]^{-1}
\end{equation}

\begin{figure}
\centering
\includegraphics[scale=.8]{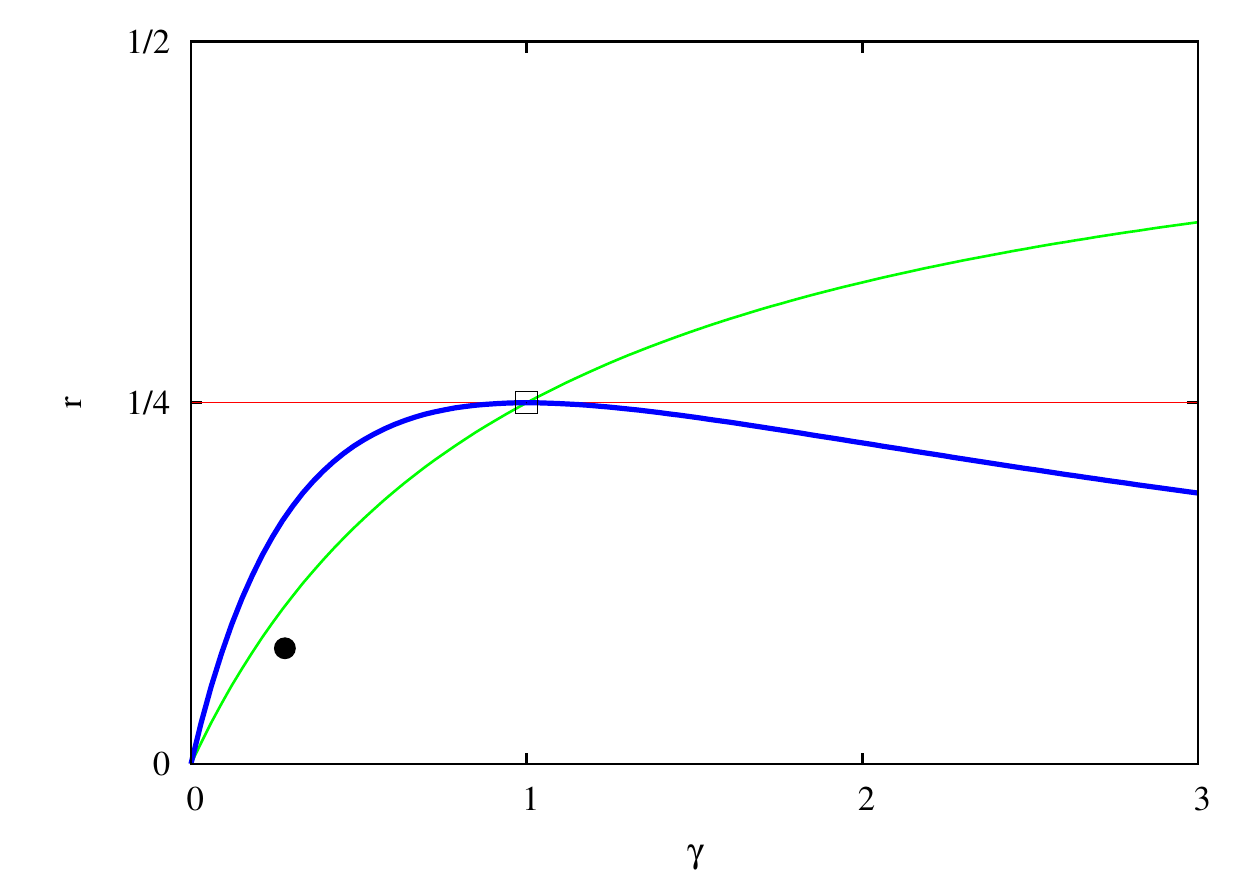}
\caption{The permissible range in the model parameter space is $0<r\leq 1/2$ and $0<\gamma<\infty$. Shown are the curves [$r=1/4$, $r=\gamma/(1+\gamma)^2$ thick blue, and $\gamma=2r/(1-2r)$ thin green] where the solution has somewhat simpler forms. At the intersection of these three curves ($\gamma = 1, r=1/4$, $\square$) the solution is particularly neat (see section \ref{special}). The experimentally measured parameter values ($\gamma = 0.28, r=0.08$) for mice epidermis in \citep{clayton07} are depicted by the $\bullet$ symbol. }
\label{r-gamma}
\end{figure}

The exact solution \eqref{F-gen} for the generating function can in principle be expanded in $x$ and $y$ to yield the probability distribution $P_{m,n}$ for arbitrary $m,n$. For instance, the system is empty with probability  
\begin{equation}
\label{P00}
P_{0,0}=\Pi_0 =  1+\gamma^{-1}\,e^{-\gamma t} -
\frac{\exp \left(\hat\gamma  e^{-\gamma t}\right)}
{\rho e^{\hat\gamma}  + \mathcal{E}(0,t)}
\end{equation}
with $\rho=\gamma/(1+\gamma)$, see Eq.~\eqref{rho}. 
The expressions for the clone size distribution are simple when $n=0$, that is for the clones without post-mitotic cells. Expanding $F(x,0,t)$ in powers of $x$ and using $F(x,0,t)=\sum_{m\geq 0} x^m P_{m,0}(t)$ we obtain 
\begin{equation}
\label{Pm0}
P_{m,0} = 
\frac{\rho \exp \left(\hat\gamma  e^{-\gamma t} + \hat\gamma\right)}
{\left[ \rho e^{\hat\gamma}  + \mathcal{E}(0,t)\right]^2 }\,\, \rho^m
\left[\frac{\mathcal{E}(0,t)}
{\rho e^{\hat\gamma}  + \mathcal{E}(0,t)}\right]^{m-1}
\end{equation}
The probabilities $P_{m,n}$ quickly become very unwieldy for $n>0$.

\section{Explicit results at the special point}
\label{special}

At the special point $\gamma_c = 1, r_c=1/4$ in the parameter space we can solve everything explicitly. Indeed, in this case $\hat\gamma =0$ and \eqref{F-gen} becomes
\begin{equation}
\label{F-simple}
F(x,y,t) = 1+(1-y)\,e^{- t} + 
\frac{1}{(x +y -2)^{-1} - t/4}
\end{equation}

Let us first extract the reduced distribution. Writing $x=z, y=z$ and noting that
\begin{equation}
\label{sumgendef}
G(z,t) \equiv F(z,z,t) = \sum_{s\geq 0} z^s\,\Pi_s(t)
\end{equation}
we conclude that
\begin{equation}
G(z,t) = 1+(1-z)\,e^{- t} -\frac{4}{t}+
\frac{4}{t}\left[1+\frac{t}{2}-\frac{t}{2}\,z\right]^{-1}
\end{equation}
Expanding the latter expression in $z$ around $z=0$ we get
\begin{subequations}
\begin{align}
&   \Pi_0  = 1 + e^{-t} - \frac{4}{t+2}
\label{Pi-0}\\
&   \Pi_1  = -e^{-t} + \frac{8}{(t+2)^2}
\label{Pi-1}\\
&   \Pi_s = \frac{8}{(t+2)^2}\left[\frac{t}{t+2}\right]^{s-1}, \quad s\geq 2
 \label{Pi-s}
\end{align}
\end{subequations}
In the scaling region
\begin{equation}
\label{scaling-s}
s\to\infty, \quad t\to \infty, \quad \frac{s}{t} = {\rm finite}
\end{equation}
equation \eqref{Pi-s} acquires a scaling form
\begin{equation}
\Pi_s \simeq \frac{8}{t^2}\,e^{-2s/t}
\end{equation}
Recall that the exact expression \eqref{Pmt} for the distribution of $A$ cells also acquires an asymptotic scaling form; in the present case $r=r_c=1/4$ it is given by 
\begin{equation}
P_m(t)  = \frac{16}{(t+4)^2}\left[\frac{t}{t+4}\right]^{m-1}
\simeq  \frac{16}{t^2}\,e^{-4m/t}
\end{equation}

Generally, by expanding \eqref{F-simple}, we obtain
\begin{equation}
   P_{0,1}  = -e^{-t} + \frac{4}{(t+2)^2}
  \label{P-01}
\end{equation}
and, for $(m,n)\ne (0,0), (0,1)$, 
\begin{equation}
 P_{m,n} = \frac{4}{(t+2)^2}\left[\frac{t}{2(t+2)}\right]^{m+n-1}\binom{m+n}{m}
\label{P-mn}
\end{equation}
The probability that the system is empty is $P_{0,0}=\Pi_0$, so it is given by Eq.~\eqref{Pi-0}. 

The clone size distribution greatly simplifies at this special point due to a mapping of our two-type branching process onto a single-type critical branching process. Indeed, at $\gamma=1, r=1/4$, the process can be reformulated as 
\begin{equation}
\label{specmodel}
 \begin{tabular}{ll}
  $C \to  C C$  ~~~ & at rate $1/2$\\
  $C \to  \emptyset$ & at rate $1/2$\\
 \end{tabular}
\end{equation}
where we assign the type $A$ or $B$ to each cell independently with probability $1/2$. 
This mapping holds if also initially we have an $A$ or a $B$ cell equiprobably. If the initial cell is type $A$, then from the solution for a  single initial $B$ cell \eqref{onlyB}, and from the solution \eqref{Pmt} of \eqref{specmodel}, we recover the behavior \eqref{P-01}--\eqref{P-mn} due to the linearity of the problem.

\section{General results}
\label{general}

In section \ref{elementary} we have found an explicit, exact expression for the generating function, equation \eqref{F-gen}, which is valid on the curve \eqref{rg-constraint}. This curve misses the parameter values ($\gamma = 0.28, r=0.08$) experimentally measured in mice tail epidermis \citep{clayton07}, see Figure~\ref{r-gamma}. In different tissues the parameters will probably take different values, so it is desirable to possess a solution in the whole range of parameters, i.e. in the strip $0<r\leq 1/2$ and $0<\gamma<\infty$. Surprisingly, using the reduction of the Riccati equation to the Sturm-Liouville equation we can find a general solution.

We start with the general backward equation \eqref{backmain} which we re-write in a canonical form
\begin{equation}
\label{nonlin}
 F' = AF^2+BF+C
\end{equation}
The coefficients of the quadratic polynomial on the right-hand side are
\begin{equation}
 A=-\frac{r}{\gamma u}, ~~~ B=\frac{1}{\gamma} \left[ 1+2r\frac{1-u}{u}\right], ~~~ C=-\frac{r}{\gamma u}(1-u)^2
\end{equation}
To transform the Riccati equation \eqref{nonlin} into the Sturm-Liouville equation we perform  the standard procedure \citep{Bender}, namely we write $F(u)$ as 
\begin{equation}
 F = - \frac{z'}{A z} = -\frac{(\log z)'}{A}
\end{equation}
After this transformation, the first order nonlinear equation \eqref{nonlin} turns into a second order linear differential equation 
\begin{equation}
\label{linz}
 z'' +\alpha z' + \beta z = 0
\end{equation}
with
\begin{equation}
\begin{split}
 \alpha &= -\left( \frac{A'}{A}+B \right)=\frac{\gamma-u-2r(1-u)}{u\gamma}\\
 \beta   &= AC = \left[ \frac{r(1-u)}{u\gamma} \right]^2
 \end{split}
\end{equation}
Now in \eqref{linz} the first derivative can be cancelled by writing $z=\Phi Z$, with the condition $\Phi'=-\alpha\Phi/2$, which leads to 
\begin{equation}
\label{Phidef}
 \Phi=e^{-\frac{1}{2}\int^u \alpha(u')du'}
\end{equation}
Then \eqref{linz} becomes 
a Shr\"odinger equation for $Z(u)$
\begin{equation}
\label{finalSch}
 Z''+ \left( \frac{4r-1}{4\gamma^2} + \frac{\gamma(1-2r)-2r}{2u\gamma^2} + \frac{1}{4u^2}\right) Z = 0
\end{equation}

Equation \eqref{finalSch} resembles the Whittaker equation. Re-scaling the variable
$u$ and making changes in notations
\begin{equation}
\label{newvars}
 g=\frac{uv}{\gamma},\quad  v = \sqrt{1-4r},\quad w= \frac{\gamma(1-2r)-2r}{2\gamma v}
\end{equation}
we recast equation \eqref{finalSch} into a canonical Whittaker differential equation
\begin{equation}
\label{finalerSch}
\frac{d^2 Z}{d g^2}+ \left(-\frac{1}{4} + \frac{w}{g} + \frac{1}{4g^2}\right)  Z = 0
\end{equation}
Its solution, up to an irrelevant constant factor, is
\begin{equation}
\label{solgen}
 Z(g) = M_{w,0}(g) + CW_{w,0}(g)
\end{equation}
where $M$ and $W$ are the Whittaker functions \citep{gradshteyn}, and $C$ is a constant to be determined from the boundary conditions.

Now we have to re-express the solution of Eq.~\eqref{solgen} in terms of the  original variables. Following the steps that have been made, but backwards, we obtain
\begin{equation}
 F(u) = -\frac{[\log z(u)]'}{A} = \frac{\gamma u}{r} \left[ \log Z(u) + \log \Phi(u)  \right]'
\end{equation}
Using $[\log\Phi(u)]'=-\alpha/2$, see Eq.~\eqref{Phidef}, we get
\begin{equation}
\label{Fuwithdiff}
 F(u) = \frac{uv}{r} \cdot \frac{M_{w,0}'(g) + CW_{w,0}'(g)}{M_{w,0}(g) + CW_{w,0}(g)} 
 + 1-u+\frac{u-\gamma}{2r} 
\end{equation}
Noting that
\begin{equation}
 \begin{split}
  M_{w,0}'(g) &= \frac{(g-2w)M_{w,0}(g)+(1+2w)M_{1+w,0}(g)}{2g} \\ 
  W_{w,0}'(g) &= \frac{(g-2w)W_{w,0}(g)- 2 W_{1+w,0}(g)}{2g}
 \end{split}
\end{equation}
we simplify \eqref{Fuwithdiff} and arrive at our main result
\begin{equation}
\begin{split}
\label{genfinal}
 F &= 1-u + \frac{u(1+v)-\gamma(1+2w)}{2r}\\
 &+ \frac{\gamma}{2r} \cdot \frac{(1+2w)M_{1+w,0}(g)-2CW_{1+w,0}(g)}{M_{w,0}(g) + CW_{w,0}(g)} 
\end{split}
\end{equation}
Recall that the parameters $g,v,w$ are given by \eqref{newvars}, and $u=(1-y)e^{-\gamma t}$. The constant $C$ in Eq.~\eqref{genfinal} is determined from the initial condition, $F(u=1-y)=x$, to give
\begin{equation}
\label{constantgen}
 C = \frac{-\theta M_{w,0}(\hat g)+(1+2w) M_{1+w,0}(\hat g)}{\theta W_{w,0}(\hat g)+ 2 W_{1+w,0}(\hat g)}
\end{equation}
Here we introduced two more shorthand notations 
\begin{equation}
 \theta = 1+2w-\hat g + \frac{2r(x-y)+y-1}{\gamma}, \quad \hat g=\frac{(1-y)v}{\gamma}
\end{equation}

The distribution of the total number of cells $\Pi_s(t)$ can be obtained from $G(z,t)=F(z,z,t)$. The survival probability of the cells at time $t$ is
\begin{equation}
\label{surv}
 S(t) = 1-F(0,0,t)
\end{equation}
where $F$ is given by \eqref{genfinal} and \eqref{constantgen}. Note that in computing $F(x=0,y=0,t)$ all the parameters that contain $x$ and $y$ simplify. Setting $x=y=0$ we get 
$u=e^{-\gamma t}$, $\theta = 1+2w- (v+1)/\gamma$, and $\hat g=v/\gamma$.

\begin{figure}
\centering
\includegraphics[scale=.8]{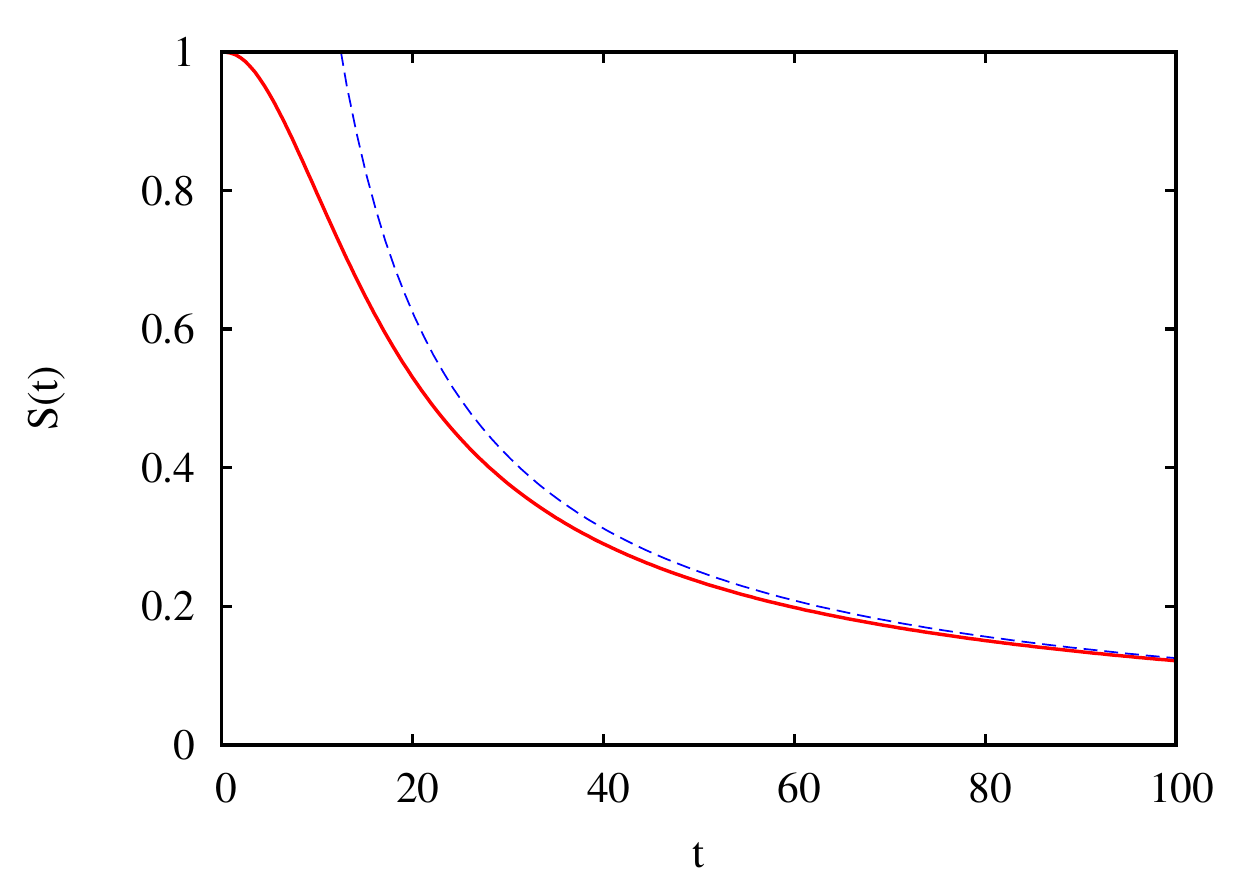}
\caption{Exact survival probability $S(t)=1-F(0,0,t)$ given by \eqref{surv}, as a function of time at the experimentally measured parameter values $\gamma = 0.28$, $r=0.08$. The dashed line is the large time asymptotic $1/rt$.}
\label{fig:survive}
\end{figure}

From the generating function \eqref{genfinal} one can easily extract the clone size distribution $P_{n,m}(t)$ or $\Pi_s(t)$ numerically. Let us start with the simpler total cell distribution $\Pi_s(t)$. The probability $\Pi_s(t)$ is the coefficient of the $z^s$ term in the power series of $G(z,t)$ as given by \eqref{sumgendef}. One way to extract $\Pi_s(t)$ is by using Cauchy's integral formula 
\begin{equation}
\label{invdef}
 \Pi_s = \frac{1}{2\pi i} \oint_C \frac{G(z)}{z^{s+1}} \, dz
\end{equation}
Here the contour $C$ goes counterclockwise around the origin in the complex $z$ plane, within the radius of convergence of $G(z)$ (we omitted the time argument for brevity). Consider a contour of a circle of radius $R$, and divide the circle into $N$ equal parts. Now the above integral \eqref{invdef} can be approximated as a sum
\begin{equation}
\label{invsum}
 \Pi_s = \frac{R^{-s}}{N} \sum_{k=0}^{N-1} G\left(Re^{ik2\pi/N}\right) e^{-iks2\pi/N} 
\end{equation}
which is the discrete Fourier transform scaled by $R^{-S}/N$. This transformation can be performed incredibly efficiently by the fast Fourier transform (FFT) method, which is implemented in most mathematical software. This method is discussed and error terms are approximated in \citep{cavers78}. Some care is needed to choose the value of $R$ to avoid numerical problems, as discussed in \citep{cavers78}. In our case the choice $R=1$ was sufficient in all examples we considered. We can check the quality of this numerical method at the special point $\gamma=1, r=1/4$, where the explicit solution for $P_{m,n}(t)$ is known \eqref{P-mn}. For example at $t=1$, with $N=32$ the numerical result for $\Pi_s(t)$ differs less than $10^{-15}$ from the exact expression for $s\le 31$, and it is precise to at least ten digits for $s\le 15$.

\begin{figure}
\centering
\includegraphics[scale=.9]{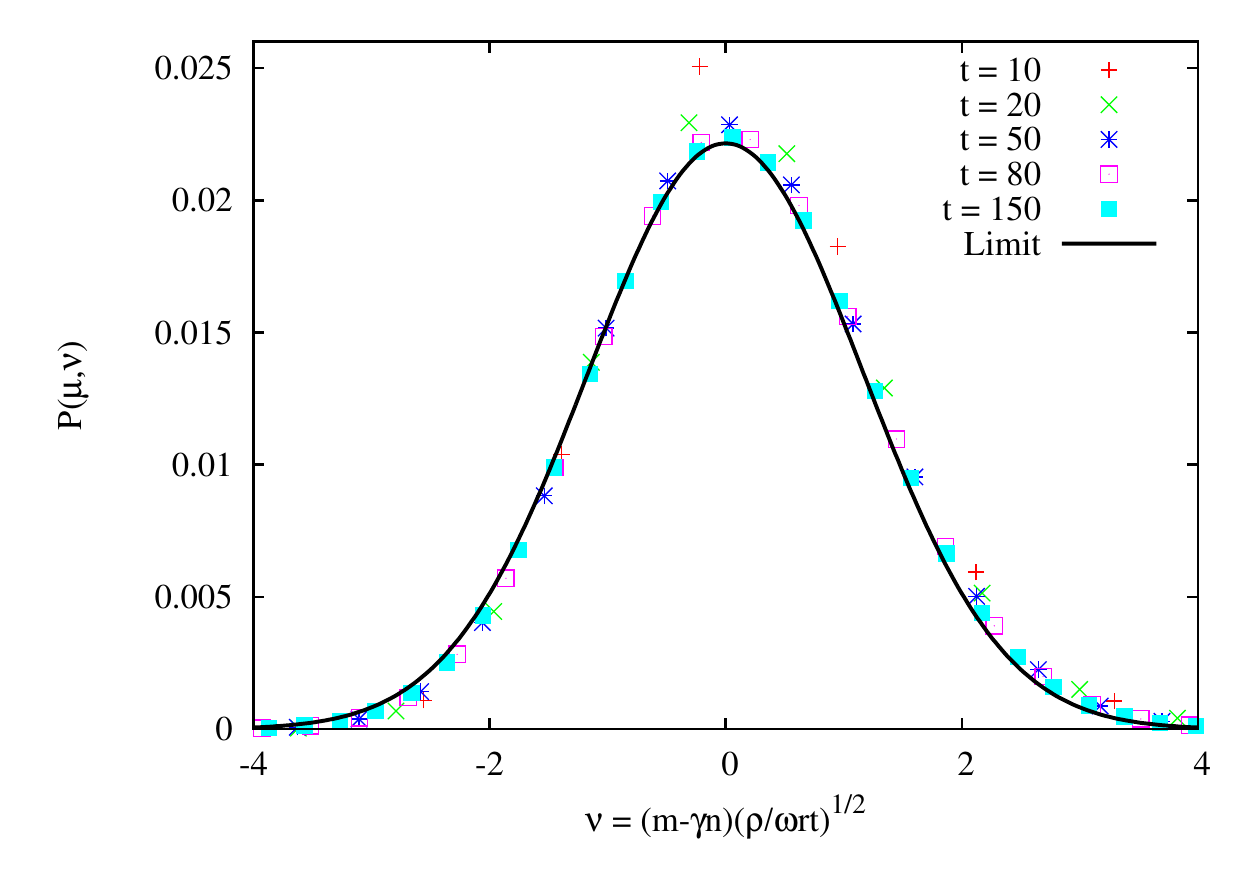}
\caption{The exact probability $P_{m,n}(t)$ of having $m$ type $A$, and $n$ type $B$ cells at time $t$, as given by \eqref{genfinal}. This probability is depicted at different times in terms of the scaling variable $\nu=(m-\gamma n)\sqrt{\rho/\omega r t}$, at $\mu=(m+n)\rho/rt=\rho/r\approx2.7$. The points collapse on the Gaussian limit curve \eqref{fullscale}.}
\label{fig:fullscale}
\end{figure}

For the full distribution $P_{m,n}(t)$ one needs two separate contour integrals in both the $x$ and the $y$ planes, which then leads to applying the discrete Fourier transform $N^2$ times. We have checked the results against the numerical solution of the forward equations \eqref{master} and found a perfect agreement (up to about 7 digits). This method has been used to obtain our figures \ref{sumscale} and \ref{fig:fullscale} for $\Pi_s(t)$ and $P_{m,n}(t)$, respectively. In \citep{clayton07} the initial cell is considered to be $A$ or $B$ with certain probabilities. The corresponding probability distribution is then a simple linear combination of the distribution $P_{m,n}(t)$ we just obtained and the trivial distribution resulting from a single initial $B$ cell \eqref{onlyB}. Since $B$ cells just die at a fixed rate, their only effect (apart from $P_{0,0}$ and  $P_{1,0}$) is to rescale $P_{m,n}(t)$.

Thus we have obtained exact results \eqref{genfinal}--\eqref{constantgen} for the generating function, which can be easily transformed back to probabilities. Moreover, these exact results simplify in a few special cases (Appendix \ref{spec_case}) and in the scaling limit (Section \ref{scaling}).

\section{Scaling limit}
\label{scaling}

In the large time limit, the distributions $P_{n,m}(t)$ and $\Pi_s(t)$ simplify. Let us consider first the reduced distribution $G(z,t)$. In the large time limit the interesting range of $s$ is $s\sim t$, see e.g.\ \eqref{scaling-s}, and therefore the interesting range of $z$ is $(1-z)\sim s^{-1}\sim t^{-1}$. Hence we consider the $t\to\infty, z\to 1$ limit with $\zeta=rt(1-z)/\rho$ kept constant, where $\rho= \gamma/(1+\gamma)$, see \eqref{rho}.

In order to perform the scaling limit we need the following small argument ($x\ll 1$) limits of the Whittaker functions
\begin{equation}
\begin{split}
 M_{w,0}(x) &= \sqrt x -wx^{3/2}+ \mathcal{O}(x^{5/2})\\
 W_{w,0}(x) &= -\sqrt x\,\, \frac{\log x + 2\gamma_E + \psi(1/2-w)}{\Gamma(1/2-w)} + \mathcal{O}(x^{3/2})
\end{split}
\end{equation}
Here $\Gamma$ is the gamma function, $\psi(z)=\Gamma'(z)/\Gamma(z)$ is the digamma function, and $\gamma_E=0.5772\dots$ is the Euler constant \citep{gradshteyn}. We also need the identity for the digamma function \citep{Bender}
\begin{equation}
 \psi\left(-\frac{1}{2}-w\right) - \psi\left(\frac{1}{2}-w\right)= \frac{2}{1+2w}
\end{equation}

\begin{figure}
\centering
\includegraphics[scale=.9]{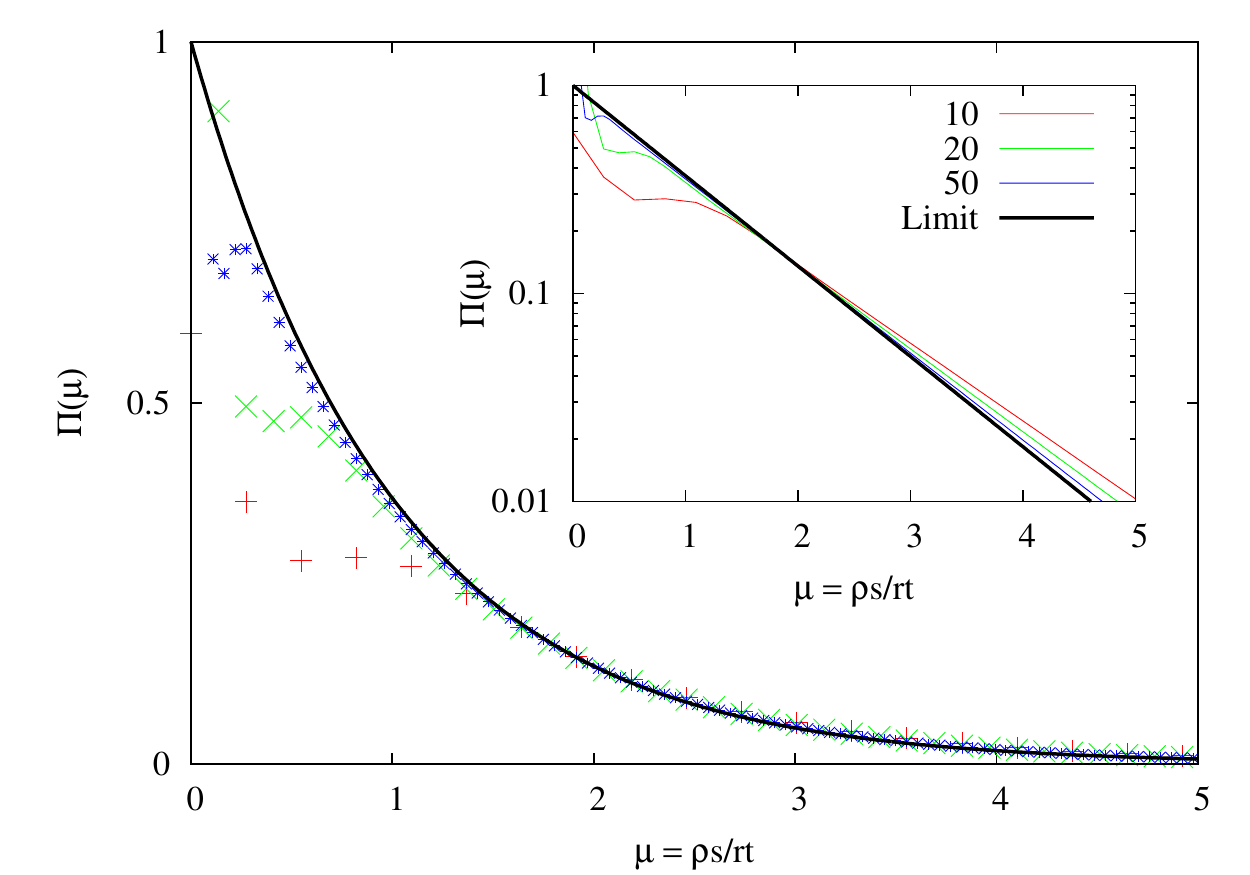}
\caption{The exact probability $\Pi_s(t)$ of having a total $s$ cells at time $t$, as given by \eqref{genfinal}. This probability is depicted at different times in terms of the scaling variable $\mu=\rho s/rt$. The points collapse on the exponential $e^{-\mu}$ limit curve \eqref{Piscale}. In the inset the same curves are re-plotted on log-scale to emphasize the tail of the distribution. The symbols are the same as in Figure~\ref{fig:fullscale}.}
\label{sumscale}
\end{figure}

Taking the $z\to 1$ limit of the constant term \eqref{constantgen} is particularly easy, since it is independent of time. We find
\begin{equation}
 C = C_0(1-z) + \mathcal{O}(1-z)^{3/2}, \quad 
 C_0 =  r \, \frac{1+\gamma}{\gamma^2}\,  \Gamma(1/2-w) 
\end{equation}
Now we substitute this expression into $F(z,z,t)$ of \eqref{genfinal}, using 
\begin{equation}
 g = \frac{v}{\gamma} u = \frac{v\zeta}{r(1+\gamma) t}\, e^{-\gamma t}
\end{equation}
Some care is needed with terms of type $W_{w,0}(g)$, where in
\begin{equation}
 \log g = -\gamma t + \log \frac{v\zeta}{r(1+\gamma) t}
\end{equation}
a term proportional to $t$ appears. In the first order of $1/t$ we obtain
\begin{equation}
\label{Gzeta}
 G(\zeta,t) 
 = 1 - \frac{1}{rt} \cdot \frac{\zeta}{\zeta+1}
\end{equation}

In the $s,t\to\infty$ scaling limit with constant $\mu=\rho s/rt$ , the generating function $G(\zeta,t)$ of \eqref{sumgendef} becomes a Laplace transform of $\Pi$
\begin{equation}
 G(\zeta,t) \to \frac{rt}{\rho} \int_0^\infty \Pi_{s(\mu)}(t) e^{-\zeta\mu} d\mu
\end{equation}
hence we can obtain the asymptotic limit of the probability $\Pi_s(t)$ by an inverse Laplace transform \citep{gradshteyn} 
\begin{equation}
\label{invlap}
 \frac{rt}{\rho}\Pi_s(t) \to  \mathcal{L}^{-1}[G(\zeta,t)] = \frac{1}{r t} e^{- \mu} + \left(1-\frac{1}{rt}\right) \delta(\mu)
\end{equation}
The first term describes the distribution of the surviving cells, while the second term stands for the extinction of cells. Consequently, the large time survival probability of the population is $1/rt$, or the extinction probability $\Pi_0(t)\sim 1- 1/rt$. In Figure \ref{fig:survive} we plotted the exact survival probability $S(t)=1-\Pi_0(t)$ of \eqref{surv} together with the large time asymptotic $1/rt$. The above asymptotic results of course agree with the explicit results in the special point of Section \ref{special}. 

From \eqref{invlap}, the regular part of the distribution $\Pi_s(t)$ can be written in a scaling form as  
\begin{equation}
\label{Pilimit}
 \Pi_s(t) = \frac{\rho}{(rt)^2} \exp\left( -\frac{\rho s}{rt} \right) = \frac{\rho}{(rt)^2} \Pi(\mu)
\end{equation}
with the time independent scaling function
\begin{equation}
\label{Piscale}
 \Pi(\mu) =  e^{-\mu} 
\end{equation}
We demonstrated this scaling in Figure~\ref{sumscale}, where the exact expressions \eqref{genfinal} for $\Pi_s(t)$ are depicted for different times as a function of the scaling variable $\mu=\rho s/rt$, and the values converge to the scaling limit \eqref{Piscale}. Note that this scaling limit has been already guessed in \citep{clayton07}, and derived in \citep{klein07} in the realm of continuous approximation, that additionally assumed that the $B$ cell population remains ``slave" to the $A$ cell population. We will see that the latter, potentially uncontrolled approximation is not merely appealing, it is asymptotically correct. This will become evident from the full distribution.

Similarly to the total cell distribution, we can also obtain the scaling limit of the whole $P_{m,n}(t)$ distribution from \eqref{genfinal}. Taking the $t\to\infty$ limit while keeping $\xi=rt(1-x)/\rho$ and $\eta=rt(1-y)/\rho$ finite, up to first order in $1/t$ we obtain
\begin{equation}
\label{fullFlim}
 F(\xi,\eta,t) = 1 - \frac{1}{rt} \cdot \frac{\xi\gamma+\eta}{\xi\gamma+\eta+1+\gamma}
\end{equation}
Of course $F(\zeta,\zeta,t)=G(\zeta,t)$ of \eqref{Gzeta} in this limit as well. Now we need to perform a double inverse Laplace transform to obtain $P_{m,n}(t)$ as a function of $m/t$ and $n/t$, in the limit $m,n,t\to\infty$. The  extinction probability is again $P_{0,0}(t)=\Pi_0(t)=1/rt$ in the first order of $1/t$. The probability $P_{m,n}(t)$ for $m,n>0$ in the scaling limit becomes
\begin{equation}
\label{sharplimit}
 P_{m,n}(t) = \frac{\gamma}{r^2 t^3}\, \exp\left[ -\frac{\rho(m+n)}{rt} \right] \delta\left( \frac{m-\gamma n}{t} \right)
\end{equation}
Hence in this limit there are precisely $\gamma$ times as many $A$ cells as $B$ cells, while the distribution of the cells is given by \eqref{Pilimit}. According to the experiments of \citet{clayton07} in skin tissue, where $\gamma=0.28$, the model predicts about four times more post mitotic cells than progenitor cells in a clone for large times.

It is possible to give a more detailed description of the cell distribution by taking a different large time limit, namely we need to take the limit $n,m,t\to\infty$ in such a way that the following fractions are finite
\begin{equation}
\label{scaling_reg}
\frac{m}{t} = O(1), \quad  \frac{n}{t} = O(1), \quad \frac{m - \gamma n}{t^{1/2}} = O(1)
\end{equation}
This limit reveals the ``shape" of the Dirac delta in \eqref{sharplimit}. This asymptotic limit is of course encoded in the exact results for the generating function \eqref{genfinal}. Unfortunately, to extract the asymptotic is far from straightforward. Indeed, even from a simple expression for the multivariate  generating function, it is usually extremely difficult to extract the asymptotic of the coefficients (let alone the exact expressions for the coefficients). This situation is perhaps surprising as in the univariate case there are various techniques, the most powerful is the use of complex analysis and the saddle point method. In the multivariate case, the usage of complex  methods is much more limited and challenging; for recent progress, see  \citep{FS09} and \citep{PW08}. In our case, there is an additional difficulty as the explicit expression for the generating function is not a simple rational function as e.g.\ in most examples in \citep{PW08}, but it involves the Whittaker functions. Hence instead of extracting the scaling limit from the exact solution, we outline another approach in the next section that also shows an independent way of handling the problem.

\section{Fokker-Planck approximation}
\label{diffuz}

We shall use a more direct procedure which is however approximate, for instance it does not even provide the asymptotically exact value, $1/rt$, that the clone size is non-zero. However, up to this amplitude one can obtain an expression for  the probability distribution $P_{m,n}(t)$ which is typically asymptotically exact in the scaling region \eqref{scaling_reg}. The method is essentially the Fokker-Planck or diffusion approximation \citep{kampen:1997aa}. One starts with the master equation \eqref{master}, and treats $m,n$ as continuous variables. This should be valid when $m,n\gg 1$. In this region one can further expand the right-hand side of \eqref{master} in the Taylor series to give (we shortly write $P$ instead of $P_{m,n}$)
\begin{equation}
\begin{split}
(m-1)P_{m-1,n} &= mP - \frac{\partial}{\partial m}\,mP
+ \frac{1}{2}\,\frac{\partial^2}{\partial m^2}\,mP+\ldots\\
mP_{m,n-1} &= mP - \frac{\partial}{\partial n}\,mP
+ \frac{1}{2}\,\frac{\partial^2}{\partial n^2}\,mP+\ldots\\
(m+1)P_{m+1,n-2} &= mP + \frac{\partial}{\partial m}\,mP 
- 2 \frac{\partial}{\partial n}\,mP
+ \frac{1}{2}\,\frac{\partial^2}{\partial m^2}\,mP 
-2\frac{\partial^2}{\partial m \partial n}\,mP + 2 \frac{\partial^2}{\partial n^2}\,mP + 
\ldots\\
(n+1)P_{m,n+1} &= nP + \frac{\partial}{\partial n}\,nP
+ \frac{1}{2}\,\frac{\partial^2}{\partial n^2}\,nP+\ldots
\end{split}
\end{equation}
Using these expansions and ignoring the higher order terms we turn the master equation into a partial differential equation 
\begin{equation}
\begin{split}
\frac{\partial P}{\partial t} &= \gamma P 
+ (\gamma - 2r - m +\gamma n)\,\frac{\partial P}{\partial n} 
+ 2r\,\frac{\partial P}{\partial m} \\
&+ rm\,\frac{\partial^2 P}{\partial m^2} - 2 rm\, \frac{\partial^2 P}{\partial m \partial n}
+ \frac{(1+2r)m + \gamma n}{2}\,\frac{\partial^2 P}{\partial n^2} 
\end{split}
\end{equation}
which is the Fokker-Planck equation  in our problem. 

Let us change $m,n$ to the variables
\begin{equation}
s= m+n, \quad \delta = m - \gamma n
\end{equation}
The Fokker-Planck equation  becomes 
\begin{equation}
\begin{split}
\label{long-FP}
\frac{\partial P}{\partial t} &= \gamma P 
+ (\gamma - \delta)\,\frac{\partial P}{\partial s} 
+ [2r+\gamma(\delta+2r-\gamma)]\,\frac{\partial P}{\partial\delta} \nonumber\\
&+ A\,\frac{\partial^2 P}{\partial s^2} 
- 2 B\, \frac{\partial^2 P}{\partial s \partial \delta}
+ C\,\frac{\partial^2 P}{\partial \delta^2} 
\end{split}
\end{equation}
with
\begin{equation}
\begin{split}
A &= \frac{2\gamma s +(1-\gamma)\delta}{2(1+\gamma)}\,,
\quad B = \gamma A, \\
C &= \gamma^2\,\frac{(1+2r)(\gamma s +\delta) + \gamma(s-\delta)}{2(1+\gamma)}
+ r(1+2\gamma)\,\frac{\gamma s +\delta}{1+\gamma}
\end{split}
\end{equation}
Since $\delta\ll s$ in the scaling region \eqref{scaling_reg}, the above coefficients simplify to 
\begin{equation}
A = \rho s, \quad B = \gamma \rho s, \quad 
C = \rho s \left[r(1+\gamma)^2+\gamma^2\right]
\end{equation}
We already know the dependence on $s$, namely 
$P\sim \exp \left(-\frac{\rho s}{rt}\right)$. To determine the dependence on $\delta$ we keep only the dominant terms in the  Fokker-Planck equation \eqref{long-FP} and obtain
\begin{equation}
\label{main}
0 = \gamma P - \gamma \delta\,\frac{\partial P}{\partial\delta} + C\,\frac{\partial^2 P}{\partial \delta^2}   
\end{equation}
Note that all terms in  \eqref{main} are of the order of $P$:
\begin{equation}
\delta\,\frac{\partial P}{\partial\delta} \sim \delta\,\frac{P}{\delta}\sim P,
\quad C\,\frac{\partial^2 P}{\partial \delta^2} \sim s \,\frac{P}{\delta^2}\sim P 
\end{equation}
The latter estimate follows from $s\sim \delta^2$ and it actually explains the choice $\delta\sim t^{1/2}$ in the scaling region \eqref{scaling_reg}. Note also that the neglected terms from  the  Fokker-Planck equation \eqref{long-FP} are indeed sub-dominant, e.g. 
\begin{equation}
A\,\frac{\partial^2 P}{\partial s^2} \sim s\,\frac{P}{s^2}=\frac{P}{s}\,,\quad
B\,\frac{\partial^2 P}{\partial s \partial \delta}\sim s\,\frac{P}{s\delta}=\frac{P}{\delta}
\end{equation}

Solving \eqref{main}, which is essentially an ordinary differential equation with respect to $\delta$, we find 
\begin{equation}
P \sim \exp \left(-\frac{\delta^2}{\omega s}\right), \quad \mbox{with}\quad  \omega = 2 (r+\rho^2)(1+\gamma)
\end{equation}
Therefore the full scaling solution reads
\begin{equation}
\label{chief}
P_{m,n}(t)  = \frac{\gamma}{(rt)^2 \sqrt{\pi\omega s}}
\exp \left( -\frac{\rho s}{rt} -\frac{\delta^2}{\omega s}\right)
\end{equation}
The amplitude, including the $1/\sqrt s$ factor, is obtained by requiring $\Pi_s=\int P(s,\delta) d\delta/(1+\gamma)$, using \eqref{Pilimit}. Note that the distribution \eqref{chief} is normalized as $\int P\, dm\,dn\,  = \int P\, ds\,d\delta\, /(1+\gamma) = (rt)^{-1}$. 

The limit distribution \eqref{chief} can be written in a scaling form
\begin{equation}
\label{fullscale}
P_{m,n}(t) = \frac{\gamma}{(rt)^{5/2}} \sqrt{\frac{\rho}{\omega}}\, P(\mu,\nu), \quad \mbox{with}\quad  
P(\mu,\nu) = \frac{e^{-\mu -\nu^2/ \mu}}{\sqrt{\pi \mu}} 
\end{equation}
with scaling variables
\begin{equation}
 \mu = \frac{\rho s}{rt} , \quad \nu = \delta \sqrt{\frac{\rho}{\omega rt}}
\end{equation}
This scaling is probed in Figure~\ref{fig:fullscale}, using exact values for $P_{m,n}(t)$ from \eqref{genfinal}. One can see that the scaling limit \eqref{chief} provides an excellent approximation already for times $t\gtrsim 10$, and the finite time curves converge to the scaling function \eqref{fullscale}. Note also that in the special point $\gamma=1, r=1/4$ the distribution \eqref{P-mn} converges exactly to the scaling limit \eqref{fullscale}.

\section{Discussion}
\label{disco}

We derived an exact solution for a two-type branching process. We investigated a specific stochastic process that has been proposed to describe measurements of murine tail epidermis  \citep{clayton07}. The chief ingredient of the stochastic process  \eqref{modeldef} is the self-duplication and differentiation of the progenitor cells without measurable contribution from stem cells. (Stem cells activate during repair from severe injuries.) The same mechanism apparently underlies the maintenance of pancreatic islets \citep{dor04} and lung homeostasis \citep{adam09}. 

An exact solution of the specific two-type branching process \eqref{modeldef} raises the hope that other two-type branching processes could be amenable to analytical treatments. Some two-type branching processes have been suggested long ago in the context of tumor formation \citep{kendall}. Indeed, cancer is often arises when a progenitor cell undergoes a series of mutations in a way that the proliferation of a mutant clone dominates the differentiation or death \citep{vog87,vog90,dingli07a,nowak06,attolini09}. The complication is that cancer typically involves multiple mutations \citep{armitage54, beer07}, so the quantitative description may require a multiple-type branching process. 

The prominent feature of our analysis is the disregard of spatial characteristics. In the context of epidermis, one might want to consider the two-dimensional version of the two-type branching process \eqref{modeldef}. The spatial model is partly amenable to analysis  \citep{klein08} due to an intimate connection with models of voting and monomer-monomer catalytic reactions \citep{PK92,LP96,Lig99}. Intriguingly, although the model presented in this paper completely disregards real space, it already provides excellent fit to experimental data.

\section*{Acknowledgments}
We are grateful for financial support from NSF grant CCF-0829541(PLK), the John Templeton Foundation, the NSF/NIH grant R01GM078986, and J. Epstein (TA).

\appendix

\section{Special Cases}
\label{spec_case}

As a check of self-consistency  it is useful to extract the explicit results of section \ref{special} from the general approach of section \ref{general}. At the special point 
$\gamma=1, r=1/4$ the potential in equation  \eqref{finalSch}  is purely quadratic $V=1/4u^2$, and the solution of \eqref{finalSch} becomes $Z=\sqrt{u} (C+\log u)$. After transforming $Z(u)$ back to $F(t)$ and fitting to the boundary conditions, we indeed recover \eqref{F-simple}.

Exact results \eqref{genfinal}--\eqref{constantgen} also simplify on a few lines in 
the parameter space. 

\subsection{Horizontal line $r=1/4$}

On this line the $u$-independent term in the potential in Eq.~\eqref{finalSch}  vanishes and the Shr\"odinger equation becomes 
\begin{equation}
\label{SE-line} 
Z'' + \frac{1}{4}\left(\frac{\gamma - 1}{\gamma^2}\,u^{-1} +u^{-2}\right)Z = 0
\end{equation}
In the $u\to 0$ limit, $Z(u)$ behaves as $\sqrt{u}$. This suggests to choose $k=\sqrt{u}$ as the basic variable and seek solution proportional to $k$. Hence we write
\begin{equation}
Z(u) = k G(k), \quad u= \frac{\gamma^2}{1-\gamma}\,k^2
\end{equation}
The amplitude $\gamma^2/(1-\gamma)$ has been chosen to get rid off $\gamma$ in the coefficients of the governing equation for $G(k)$:
\begin{equation}
\label{Bessel}
\frac{d^2 G}{d k^2} + \frac{1}{k}\,\frac{d G}{d k}  - G = 0
\end{equation}
Solutions to this equation are linear combination of the modified Bessel function $I_0(k)$ and $K_0(k)$, i.e.
\begin{equation}
G(k) = C_1 I_0(k) + C_2 K_0(k)
\end{equation}
Then the function $F(u)$ is given by 
\begin{equation}
F=1+u+ 2\gamma k\,\,
\frac{C I_1(k) -  K_1(k)}{C I_0(k) +  K_0(k)}
\end{equation}
where we used identities $I_0'(x)=I_1(x), K_0'(x)=-K_1(x)$ and $C=C_1/C_2$. 
We can re-write this as
\begin{equation}
F(x,y,t)=1+(1-y)e^{-\gamma t}+ 2\gamma k\,\,
\frac{C I_1(k) -  K_1(k)}{C I_0(k) +  K_0(k)}
\end{equation}
with
\begin{equation}
\label{k-def}
k=\gamma^{-1}e^{-\gamma t/2}\sqrt{(1-\gamma)(1-y)}
\end{equation}
where $C=C(x,y)$ is determined by matching to the initial condition $F(x,y,t=0)=x$. One gets 
\begin{equation}
C = \frac{\gamma\kappa K_1(\kappa) - \left(1-\frac{x+y}{2}\right)K_0(\kappa)}
{\gamma\kappa I_1(\kappa) + \left(1-\frac{x+y}{2}\right)I_0(\kappa)}
\end{equation}
with
\begin{equation}
\kappa = k(t=0)=\gamma^{-1}\sqrt{(1-\gamma)(1-y)}
\end{equation}

\subsection{Special curve $\gamma=2r/(1-2r)$}

On this curve $w=0$, hence the term proportional to $g^{-1}$ vanishes in the Schr\"odinger equation \eqref{finalerSch}, which then can be solved in terms of Bessel functions. Here instead, we derive the simplified form from the general solution \eqref{finalerSch}. For this we need some limit properties \citep{gradshteyn} of the Whittaker functions 
\begin{equation}
 \begin{split}
   M_{0,0}(g) &= \sqrt{g} I_0 (g/2)\\
   W_{0,0}(g) &= \sqrt{g/\pi}\,  K_0(g/2)\\
   M_{1,0}(g) &= (1-g)\sqrt{g} I_0(g/2) + g^{3/2} I_1(g/2) \\
   W_{1,0}(g) &= \frac{1}{2\sqrt \pi} \left[ -(1-g)\sqrt{g} K_0(g/2) + g^{3/2} K_1(g/2) \right]\\
 \end{split}
\end{equation}
By using these expressions in \eqref{genfinal}, we obtain
\begin{equation}
 F(u) = 1-u+\frac{u}{2r}\left[ 1 + v \cdot \frac{ I_1(g/2) - CK_1(g/2)}{ I_0(g/2) + CK_0(g/2)} \right]
\end{equation}
with constant 
\begin{equation}
 C = \frac{-\chi I_0(\hat g/2) + v I_1(\hat g/2)}{\chi K_0(\hat g/2)+ v K_1(\hat g/2)}, \quad \chi = \frac{2r(x-y)}{1-y}-1
\end{equation}




\end{document}